\documentclass[journal]{IEEEtran}
\usepackage{cite}
\ifCLASSINFOpdf
\else
\fi
\usepackage[cmex10]{amsmath}
\usepackage{array}
\usepackage{mdwmath}
\usepackage{mdwtab}
\usepackage{graphicx}
\usepackage{euscript}
\usepackage{amsfonts}
\usepackage{bigstrut}
\usepackage{tabularx}
\usepackage{cite}
\usepackage{amsmath}
\usepackage{algorithm}
\usepackage[noend]{algpseudocode}
\makeatletter
\def\BState{\State\hskip-\ALG@thistlm}
\makeatother
\usepackage{enumitem}
\usepackage{upgreek}
\usepackage{dsfont}
\usepackage{amsmath}
\usepackage{amsthm}
\usepackage{amssymb}
\usepackage{xcolor}
\usepackage{mathtools}

\usepackage{hyperref}
\usepackage{epstopdf} 
\newcommand\numeq[1]%
{\stackrel{\scriptscriptstyle(\mkern-1.5mu#1\mkern-1.5mu)}{=}}
\usepackage{bbm}
\DeclareMathOperator{\acos}{acos}
\DeclareMathOperator{\asin}{asin}

\DeclareMathOperator{\acot}{acot}

\DeclareMathOperator{\sinc}{sinc}
\DeclareMathOperator*{\argmin}{arg\,min}

\newcommand\tinyEarth{\vcenter{\hbox{\scalebox{0.5}{$\oplus$}}}}
\newcommand\LoS{\vcenter{\hbox{\scalebox{0.5}{$\mathrm{LoS}$}}}}
\newcommand\NLoS{\vcenter{\hbox{\scalebox{0.5}{$\mathrm{NLoS}$}}}}

\begin{document}

\title{Modeling Uplink Coverage Performance in Hybrid Satellite-Terrestrial Networks}

\author{Bassel Al Homssi,~\IEEEmembership{Member,~IEEE}, and Akram~Al-Hourani,~\IEEEmembership{Senior Member,~IEEE}.
	
\thanks{B. Al Homssi and A. Al-Hourani are with the School of Engineering, RMIT University, Melbourne, Australia. E-mail: bhomssi@ieee.org and akram.hourani@rmit.edu.au.}
}

\maketitle

\begin{abstract}
Once deemed a far-fetched vision, emerging deployments of massive satellite constellations will soon offer true global coverage. When these constellations overlay the evolving terrestrial networks, a new hybrid continuum is formed which has the potential to provide uninterrupted coverage. In this paper, we provide an analytic framework for the uplink coverage probability in hybrid satellite-terrestrial networks. The framework extends well-developed terrestrial models that utilize tools from stochastic geometry by incorporating an additional layer that fits the emerging next generation satellite-terrestrial networks. The paper captures the impact of both (i) the constellation size, and (ii) the terrestrial base station density on the coverage of the uplink traffic which is dominant in applications relying on wireless sensor networks, such as the Internet of Things. This framework provides insights that guide the design of hybrid network infrastructures for a desired quality of service.
\end{abstract}


\IEEEpeerreviewmaketitle 

\section{Introduction}
\begin{figure}
	\centering
	\includegraphics[width=0.9\linewidth]{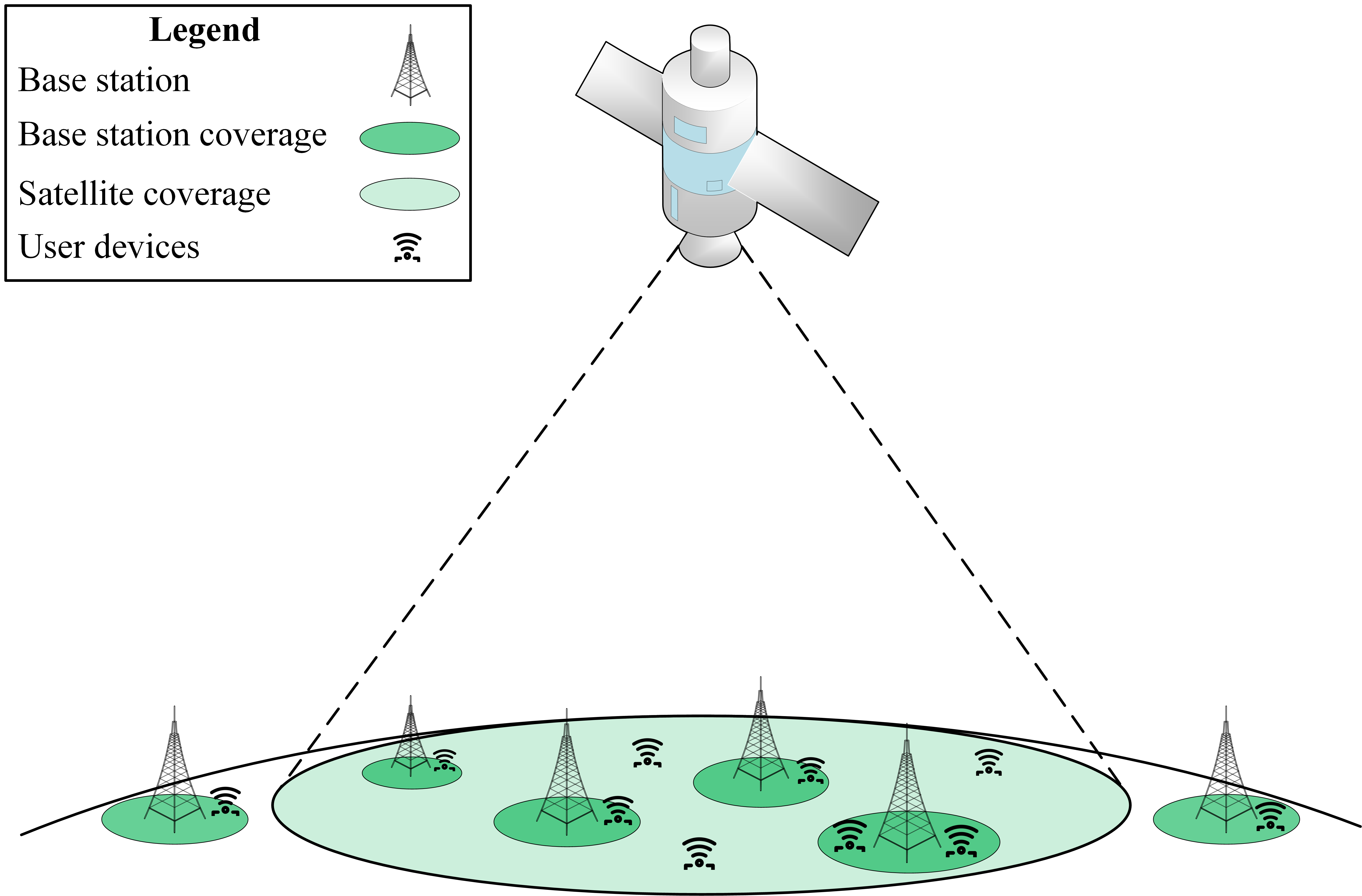}
	\caption{An illustration of the hybrid satellite-terrestrial continuum, where terrestrial gaps are bridged by overlaying satellite coverage.}
	\label{Fig_Abstract}
	\footnotesize
\end{figure}

\IEEEPARstart{N}{ext} generation terrestrial networks are expected to provide improved mobility, connectivity, and capacity to rival the current soar in demand for higher quality and faster uplink communications, especially with the deployment of massive wireless Internet-of-Things (IoT) sensor networks. A large number of these IoT sensors are expected to be dispersed in remote locations serving various applications such as smart agriculture, asset tracking, mining, oil rigs, and environmental sensing. However, remote and off-shore deployments are not typically supported by terrestrial coverage due to technical difficulties and cost limitations. Therefore, it is imperative to have a more flexible network architecture that incorporates both terrestrial and satellite technologies to facilitate seamless coverage continuum. Driven by this vision, the third generation partnership project (3GPP) is actively looking into hybrid satellite-terrestrial networks under the 5G framework~\cite{7811844}. The satellite link can provide complementary coverage to the underlying terrestrial infrastructure and ensure service continuity. Fig.~\ref{Fig_Abstract} depicts the concept of hybrid satellite-terrestrial network where a seamless coverage is potentially attained~\cite{1522108}.
The downside is that unlike terrestrial base stations (BS), satellites are usually very expensive and require long lead time to deploy. This is only exacerbated with the need for massive numbers of satellites required to achieve adequate global coverage. As a result, network planning is of paramount importance to sustain cost-effective deployment which meets operator targets. Analytic approaches can usually provide deeper insights into the different parameters impacting the performance of the complex satellite constellation networks. Tools from stochastic geometry are widely deployed to capture the performance of terrestrial networks, these tools have been recently extended to analyze the behavior of satellite networks~\cite{9313025}. Example of recent work can be found in~\cite{9079921,9347980} for analyzing the coverage probability by relying on the log-distance path-loss channel model typically used to characterize terrestrial links. However, the vast proportion of the satellite communication occurs under free-space conditions and only interacts with the clutter layer near ground~\cite{9313025,9257490}. Moreover, the literature lacks an analytic framework that provides the uplink coverage for hybrid satellite-terrestrial networks. Thus, in this letter we provide an analytic framework that models the uplink coverage probability for hybrid satellite-terrestrial networks in rural environments using satellite specific path-loss and shadowing model. The hybrid coverage is assumed to be modeled using two independent and homogeneously distributed sets of receivers; (i) a satellite constellation, and (ii) terrestrial BS. Accordingly, the framework presents a relationship between the coverage probability and the operation parameters for both networks, including; (i) the number of satellites, (ii) base station (BS) density, and (iii) ground devices density. The framework also presents an expansion strategy based on a given target quality-of-service (QoS) and the projected devices increase. The contributions of this work are summarized as follows:
\begin{itemize}
    \item It presents an analytic approach for the uplink coverage probability in satellite-terrestrial networks suitable for IoT over satellite applications.
    \item It derives analytic relations between the number of satellites and BS density for a prescribed QoS.
    \item It depicts an expansion strategy for satellite-terrestrial networks in terms of satellite numbers and BS density to cater for the projected increase in ground users. 
\end{itemize}

\section{System Model}
\begin{figure}
	\centering
	\includegraphics[width=0.8\linewidth]{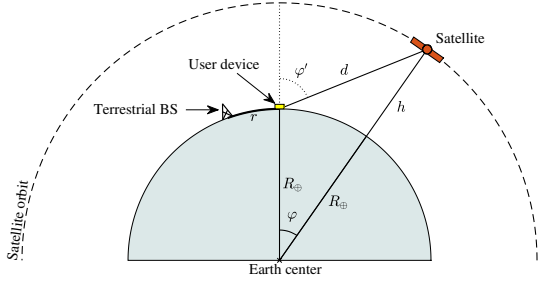}
	\caption{A user-centric simplified representation of the hybrid network where the user is jointly being served by a terrestrial BS and a satellite.}
	\label{Fig_GeometricModel}
	\footnotesize
\end{figure}
\subsection{Geometric Model}
\noindent
We define three independent point processes that respectively represent the locations of (i) the satellite constellation, (ii) terrestrial BS, and (iii) users.
The satellite locations are considered to be randomly and homogeneously scattered on a constellation sphere of altitude $h$ above the mean sea level, each satellite follows a random circular orbit with the same radius. This assumption preserves the uncorrelation and uniform distribution overtime. We show later that this assumption provides the upper bound and behaves closely in comparison to deterministic Walker satellite constellations. Therefore, the satellite locations can be approximated by the homogeneous Poisson point process (PPP) as shown in~\cite{9313025}. Consider a constellation is composed of $N_{\mathrm{s}}$ satellites orbiting on a spherical surface with an altitude of $h$, then the average satellite density is given as,
\begin{equation}
	\lambda_{\mathrm{s}} = \frac{N_{\text{s}}}{4\pi (R_{\tinyEarth} + h)^2} ,
\end{equation}
where $R_{\tinyEarth}$ is the Earth's average radius. Each user is assumed to connect to the satellite that is closest to its local zenith due to better radio propagation conditions. Using geometric reasoning, we represent the user-centered zenith angle $\varphi^\prime$ in terms of the Earth-centered zenith angle $\varphi$, as follows,
\begin{equation}
    \varphi^\prime=\acot\left(\frac{\cos\varphi-\alpha}{\sin\varphi}\right),
\end{equation}
where $\alpha = R_{\tinyEarth}/(R_{\tinyEarth} + h)$, and both angles are illustrated in Fig.~\ref{Fig_GeometricModel}. Accordingly, the probability density function (PDF) of the contact angle $\varphi_\mathrm{o}$ is obtained in~\cite{9313025} as follows,
\begin{equation}
	f_{\varphi_\mathrm{o}}(\varphi) = \frac{N_\mathrm{s}}{2}\sin\varphi\exp\left(-\frac{N_\mathrm{s}}{2}[1-\cos\varphi]\right) .
\end{equation}
On the other hand, the terrestrial BS locations are represented with an independent and homogeneous PPP on Earth's surface $\Phi_\mathrm{b}$ on $\mathbb{S}^2$ with density ${\lambda_{\mathrm{b}}}$ where $\mathbb{S}^2\subset\mathbb{R}^3$  represents Earth's spherical surface. In this model, we assume that each device connects to its nearest BS. Since both the devices and BS are located on the Earth's surface, the distances between the users and their serving BS are very small, in comparison to the Earth's circumference, therefore we can approximate the Earth surface as flat and infinite from the user perspective. As a result, the contact-distance is simplified to the well-known 2-dimensional PPP model, where the the PDF of the contact distance $R_\mathrm{o}$ is given by~\cite{HaenggiBook},
\begin{equation}
	f_{R_{\mathrm{o}}}(r) = 2\pi \lambda_{\mathrm{b}}r \exp(-\pi\lambda_{\mathrm{b}}r^2),
\end{equation}
where $r$ is the distance between the user and its serving BS.
Moreover, the users are assumed to be randomly and homogeneously dispersed on the Earth's surface. Hence, we represent their locations with a third independent PPP on Earth's surface, i.e. $\Phi_\mathrm{d}$ on $\mathbb{S}^2$ with density $\lambda_{\mathrm{d}}$. In this proposed framework we assume that both the satellite constellation and terrestrial BS listen to the frames transmitted by users. As such, when a user transmits a frame it is suffice for either of the networks to receive it for the transmission to be successful. Fig~\ref{Fig_GeometricModel} illustrates a user-centric representation of the hybrid network showing the main geometric parameters.
\subsection{Channel Model}
\noindent
We assume that users are equipped with hemispheric-gain antennas (pointed upward) such that the uplink EIRP is uniform from the satellites' perspective, denoted as $P$. We have two channel models in a hybrid network; (i) ground-to-satellite and (ii) user-to-terrestrial BS.
For the Satellite-to-Ground channel, we follow the empirical model obtained from our previous work in~\cite{9257490}, in which the received power at the satellite is modeled as,
\begin{equation}\label{Eq_ReceivedPower}
	S_\mathrm{s} = P~l~\zeta ,
\end{equation}
where $l$ is the free-space path-gain obtained as follows,
\begin{equation}
	l(\varphi) = \frac{l_\mathrm{o}l_{\mathrm{air}}}{d^{2}} =  \frac{l_{\mathrm{o}}l_{\mathrm{air}}}{R_{\tinyEarth}^2 + (R_{\tinyEarth} + h)^2 -2R_{\tinyEarth}(R_{\tinyEarth}+h)\cos\varphi} ,
\end{equation}
where $l_\mathrm{air}$ is the air absorption attenuation caused by the resonance of gas and water vapor between Earth's surface and the satellite, $l_\mathrm{o}$ is the path-gain constant ${l_\mathrm{o} = c^2/(4\pi f)^{2}}$ where $c$ is the speed of light and $f$ is the center carrier frequency. $\zeta$ in \eqref{Eq_ReceivedPower} is the excess path-gain, i.e. the reciprocal of the excess path-loss. The distribution of the excess path-gain follows a mixed Gaussian distribution as follows~\cite{9257490},
\begin{equation}
	\zeta \mathrm{[dB]} \sim p_{\LoS}\mathcal{N}(-\mu_{\LoS} ,\sigma_{\LoS}^2) + p_{\NLoS}\mathcal{N}(-\mu_{\NLoS} ,\sigma_{\NLoS}^2) ,
\end{equation}
where $\mu_{\LoS},\sigma_{\LoS},\mu_{\NLoS},\sigma_{\NLoS}$ depend on the propagation environment, $p_{\LoS}$ and $p_{\NLoS}$ are the line-of-sight (LoS) and  non-line-of-sight (NLoS) probabilities respectively. The LoS probability is derived as a function of the contact angle as follows,
\begin{equation}
	p_{\LoS}=\exp(-\beta \cot\theta)=\exp\left(-\frac{\beta \sin\varphi}{\cos\varphi - \alpha}\right) ,
\end{equation}
where $\theta$ is the ground user elevation angle and $\beta$ is a parameter that depends on the propagation environment.
On the other hand, for the terrestrial channel model we utilize the common log-distance path-loss model, where the received power at the terrestrial BS is given as follows,
\begin{equation}
	S_\mathrm{b} = Pg~b~l_{\mathrm{o}}r^{-a} ,
\end{equation}
where $a$ is the path-loss exponent, $b$ is a model constant, and $g$ is the fading process assumed to follow a Rayleigh model.

\section{Uplink Interference}
\begin{figure}
	\centering
	\includegraphics[width=0.8\linewidth]{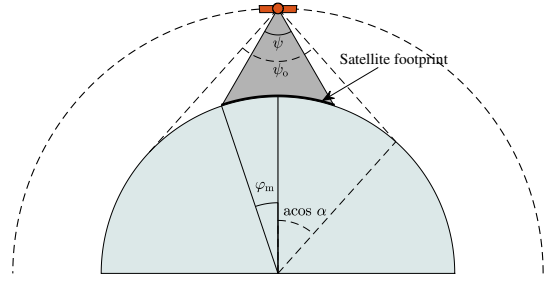}
	\caption{A satellite-centric representation of the beamwidth effect on its footprint and its maximum Earth-centered zenith angle.}
	\label{Fig_SatAngle}
	\footnotesize
\end{figure}

Wireless communications rely on the reuse of spectral resources to achieve better spectral efficiency, thus networks utilize well-designed \emph{access systems} that enable the coexistence of different transmissions. Such access systems endeavor to mitigate the impact of co-channel interference by intelligently scheduling spectral resources. We capture the effect of access systems in mitigating the interference via a factor ${\kappa\in[0,1]}$. This factor acts on the total interference power, where a value of ${\kappa=0}$ indicates an ideal system, i.e. full interference mitigation, and a value of ${\kappa=1}$ indicates the worst-case scenario, i.e. no interference mitigation. For a generic network, the effect of the access system in the satellite link is denoted as, $\kappa_\mathrm{s}$ and that of the terrestrial link is denoted as, $\kappa_\mathrm{b}$, where two types of interference are formulated in a hybrid system; (i) at the satellite and (ii) at the BS receivers. At a given time instance, a satellite can only serve a certain region underneath it, named the \textit{satellite footprint} as depicted in Fig~\ref{Fig_SatAngle}, this region is  either confined because of the antenna beamwidth or the Earth's curvature. We denoted this spherical cap as $\mathcal{A}$ having an Earth-centered apex angle of $2\varphi_{\mathrm{m}}$ and radius $R_{\tinyEarth}$, where $\varphi_{\mathrm{m}}$ is the maximum Earth-centered zenith angle which is given by,
\begin{equation}
	\varphi_\mathrm{m} =
	\begin{cases}
		\asin\left(\frac{1}{\alpha}\sin \frac{\psi}{2}\right) - \frac{\psi}{2} ,& \psi<2\asin\alpha \\
		\acos \alpha ,& \psi\geq2\asin\alpha\\
	\end{cases}~,
\end{equation}
where $\psi$ is the satellite ideal beamwidth. Fig~\ref{Fig_SatAngle} also illustrates the effect of the beamwidth on $\varphi_\mathrm{m}$. From the satellite perspective, the total interference power acting on the signal transmitted by a user $x_\mathrm{o}$ is given by,
\begin{equation}
	I_{\mathrm{s}} = \sum_{x_i\in \Phi_\mathrm{d}\cap\mathcal{A}\backslash x_o} \kappa_\mathrm{s} P l_i \zeta_i~.
\end{equation}
When considering the large swath of Earth covered by the satellite, the practical number of interfering devices converges to the mean value, $\lambda_\mathrm{d} ||\mathcal{A}||$ if all devices are assumed to be active. Accordingly, the random variations in the interference power are small and thus could be ignored. As a result, it is sufficient to describe the interference by its average power to understand the performance of the system. By invoking Campbell's theorem of sums~\cite{HaenggiBook}, we obtain the average interference  power as follows,
\begin{align}\label{Eq_12}
	\bar{I_\mathrm{s}} =  \mathbb{E}\left[I_\mathrm{s}\right] &=\mathbb{E}_\zeta\left[2\pi R_{\tinyEarth}^2D\lambda_{\mathrm{d}}\int_{0}^{\varphi_{\text{m}}} \kappa_\mathrm{s}P l(\varphi)\zeta(\varphi)\sin\varphi\mathrm{d}\varphi\right] \nonumber\\ &\stackrel{(\mathrm{a})}{=} 2\pi R_{\tinyEarth}^2D\lambda_{\mathrm{d}}P \int_{0}^{\varphi_{\text{m}}} \kappa_\mathrm{s} l(\varphi)\bar{\zeta}(\varphi)\sin\varphi\mathrm{d}\varphi ,
\end{align}
where the average density of active users is a homogeneously thinned process, $D\lambda_\mathrm{d}$ resulting from the spatial duty cycle denoted as $D$ and step (a) results from the fact that the excess path-gain $\zeta$ is independent from the geometric process. The term, $2\pi R_{\tinyEarth}^2 \sin\varphi~\mathrm{d}\varphi$ in \eqref{Eq_12}, represents the circular strip area on Earth's spherical surface for a given zenith angle. Since $\zeta$ is modeled as Gaussian mixture model (GMM), its mean value is derived as follows,
\begin{equation}
	\bar{\zeta} = p_{\LoS}\exp\left(\frac{\rho^2\sigma_{\LoS}^2}{2}-\rho\mu_{\LoS}\right)+ p_{\NLoS}\exp\left(\frac{\rho^2\sigma_{\NLoS}^2}{2}-\rho\mu_{\NLoS}\right) ,
\end{equation}
where $\rho = \ln 10/10$. On the other hand, from the terrestrial BS perspective, the aggregated interference power acting on the signal transmitted by a user $x_\mathrm{o}$ is given by,
\begin{equation}
	I_{\mathrm{b}} = \sum_{x_i\in \Phi_\mathrm{d}\backslash x_o} \kappa_\mathrm{b} P h_i r^{-a}~.
\end{equation}
If the path-loss exponent is relatively large, then the interference is significantly influenced by variations in the number of nearby devices. As such we deal with the terrestrial interference as a random variable. The expression of the interference's Laplace Transform (LT) is derived based on~\cite{HaenggiBook} as follows,
\begin{equation}
	\mathcal{L}_{I_\mathrm{b}}(s) = \exp\left[- \pi D\lambda_{\mathrm{d}}\frac{(\kappa_\mathrm{b}Pbl_\mathrm{o}s)^\frac{2}{a}}{\sinc\frac{2}{a}}\right] ,
\end{equation}
where the LT of $I$ is defined as ${\mathcal{L}_I(s) = \mathbb{E}[\exp(-sI)]}$.
\section{Hybrid Coverage Probability}
In the proposed hybrid network model, it is suffice to receive the transmission from a user by either the satellite constellation or the terrestrial BS. Thus, a transmission will only fail if both of these two conditions are satisfied: (i) the signal-to-interference-and-noise ratio (SINR) of the satellite link $\gamma_\mathrm{s}$ is below the service threshold  $\gamma_\mathrm{o}$, \textit{and} (ii) the SINR of the terrestrial link is also below $\gamma_\mathrm{o}$. Accordingly, the success probability of the hybrid coverage for a target SINR is formulated as follows, 
\begin{align}\label{Eq_hybridCov}
    \bar{p}_\mathrm{c} &=  1 - \mathbb{E}\left[\mathbb{P}(\gamma_{\mathrm{b}}<\gamma_\mathrm{o},\gamma_{\mathrm{s}}<\gamma_\mathrm{o})\right]\nonumber\\
	&\stackrel{(\mathrm{a})}{=} 1 - \mathbb{E}\left[\mathbb{P}(\gamma_{\mathrm{s}}<\gamma_\mathrm{o})\right]~\mathbb{E}\left[\gamma_{\mathrm{b}}<\gamma_\mathrm{o})\right] \nonumber\\
	&=1 - \left[1-\bar{p}_\mathrm{s}(N_\mathrm{s})\right]~\left[1-\bar{p}_\mathrm{b}(\lambda_\mathrm{b})\right]~,
\end{align}
where step (a) stems from the fact that both the channel and geometric processes of the satellite and terrestrial networks are mutually independent.
\begin{figure}
	\centering
	\includegraphics[width=\linewidth]{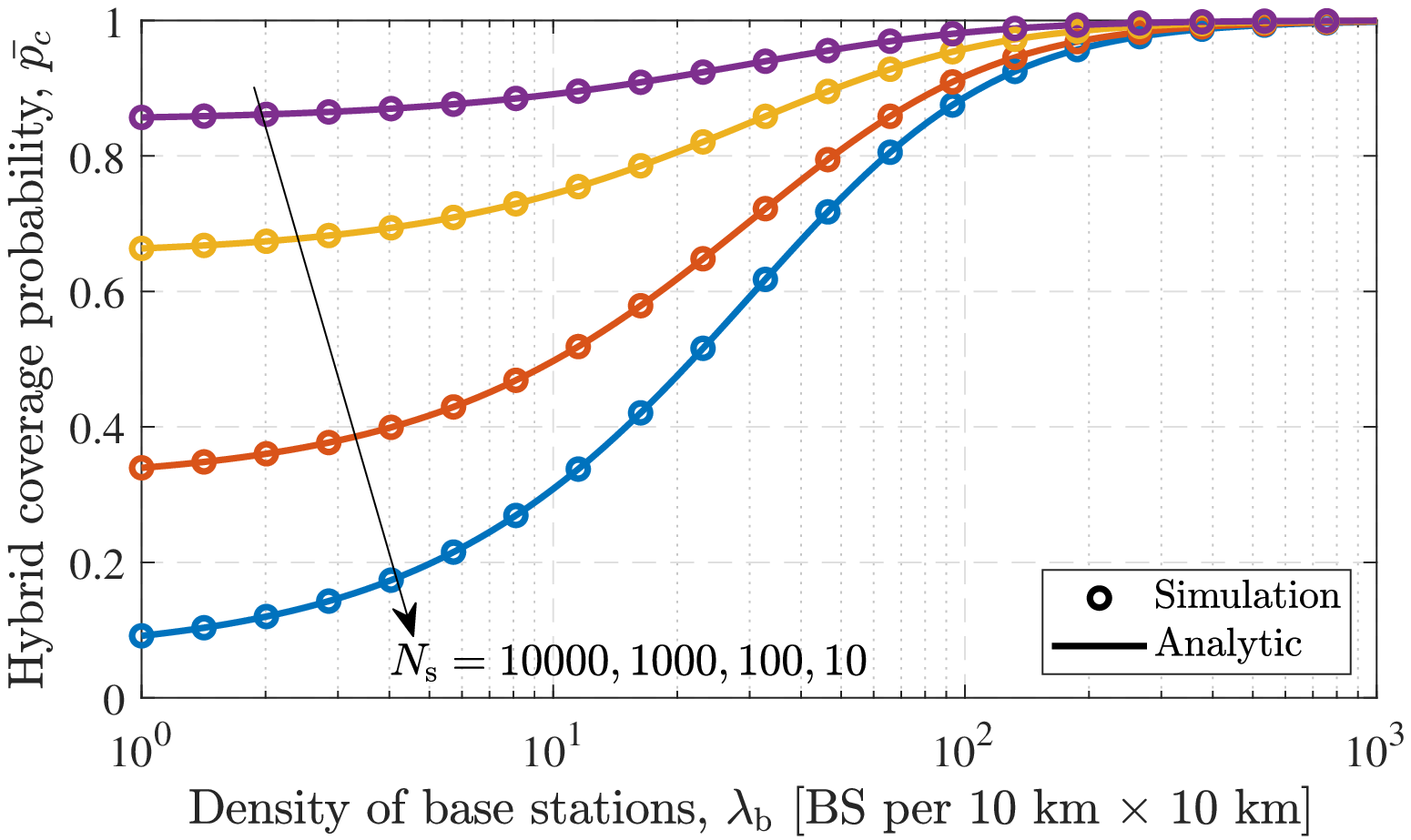}
	\caption{Average hybrid probability of coverage for variable $\lambda_{\mathrm{b}}$ showing the impact of the constellation size $N_\mathrm{s}$, comparing analytic and simulation results.}
	\label{Fig_ProbBS}
	\footnotesize
\end{figure}
\begin{figure}
	\centering
	\includegraphics[width=\linewidth]{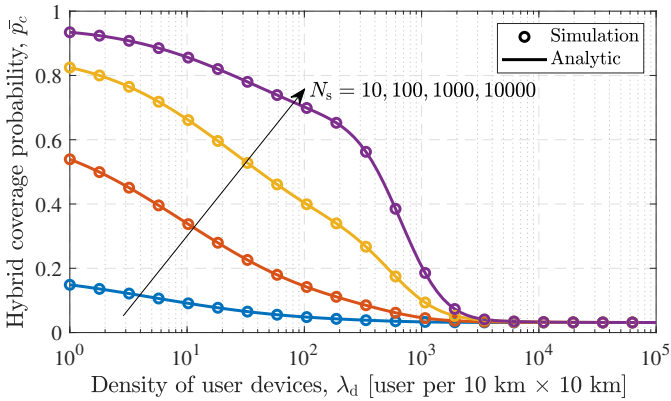}
	\caption{Average hybrid coverage probability for variable $\lambda_{\mathrm{d}}$ showing the impact of the constellation size $N_\mathrm{s}$, comparing analytic and simulation results.}
	\label{Fig_ProbIoT}
	\footnotesize
\end{figure}
The SINR of the received signal at the serving satellite is defined as,
$\gamma_\mathrm{s} = S_\mathrm{s}/(I_\mathrm{s} + W_\mathrm{s})$, where $W_\mathrm{s}$ is the average noise power observed at the satellite receiver. The resulting satellite coverage probability is thus derived as,
\begin{align}\label{Eq_SatCov}
    \bar{p}_\mathrm{s}(N_\mathrm{s}) &= \mathbb{E}_{\varphi_\mathrm{o}}\left[\mathbb{P}\left(\zeta > \frac{\gamma_{\mathrm{o}}[\bar{I}_\mathrm{s} + W_{\mathrm{s}}]}{P l(\varphi_\mathrm{o})}\right)\right] \nonumber\\ &= \mathbb{E}_{\varphi_\mathrm{o}}\left[1 - F_{\zeta}\left( \frac{\gamma_{\mathrm{o}}[\bar{I}_\mathrm{s} + W_{\mathrm{s}}]}{P l(\varphi_\mathrm{o})}\right)\right] \nonumber\\ &=  \int_{0}^{\varphi_{\text{m}}}\left[1-F_\zeta\left(\frac{\gamma_{\mathrm{o}}[\bar{I}_\mathrm{s} + W_{\mathrm{s}}]}{P l(\varphi)}\right)\right]f_{\varphi_\mathrm{o}}(\varphi)\mathrm{d}\varphi ,
\end{align}
where $F_\zeta(.)$ is the CDF of the mixed GMM path-gain random variable (in linear form) is obtained as follows,
\begin{align}\label{Eq_Fzeta}
    F_\zeta(x) = \frac{1}{2} +  &\frac{p_{\LoS}}{2}\text{erf}\left[\frac{10\log_{10} x + \mu_{\LoS}}{\sqrt{2}\sigma_{\LoS}}\right]\nonumber\\ &+ \frac{p_{\NLoS}}{2}\text{erf}\left[\frac{10\log_{10} x + \mu_{\NLoS}}{\sqrt{2}\sigma_{\NLoS}}\right] .
\end{align}
In regards to the terrestrial network, the SINR of the received signal at the serving BS is defined as, ${\gamma_\mathrm{b} = S_\mathrm{b}/(I_\mathrm{b} + W_\mathrm{b})}$, where $W_\mathrm{b}$ is the terrestrial BS's average noise power. The resulting terrestrial coverage probability is derived as,
\begin{align}\label{Eq_TerCov}
    \bar{p}_\mathrm{b}&(\lambda_\mathrm{b}) = \mathbb{E}_{R_\mathrm{o}}\left[\mathbb{P}\left(g>\frac{\gamma_\mathrm{o}r^a(I_\mathrm{b} + W_\mathrm{b})}{Pbl_\mathrm{o}}\right)\right] \nonumber\\
    &= \mathbb{E}_{R_\mathrm{o}}\left[\mathbb{E}_{I_\mathrm{b}}\left[\exp\left(-\frac{\gamma_\mathrm{o}r^aI_\mathrm{b}}{Pbl_\mathrm{o}}\right)\right]\exp\left(-\frac{\gamma_\mathrm{o}r^aW_\mathrm{b}}{Pbl_\mathrm{o}}\right)\right] \nonumber\\
    &=\int_{0}^{\infty}\mathcal{L}_{I_\mathrm{b}}\left(\frac{\gamma_{\mathrm{o}}r^a}{Pbl_\mathrm{o}}\right)\exp\left(-\frac{\gamma_{\mathrm{o}}r^aW_{\mathrm{b}}}{Pbl_\mathrm{o}}\right)f_{R_{\mathrm{o}}}(r)\mathrm{d}r.
\end{align}
\noindent
We depict in Fig.~\ref{Fig_ProbBS} the coverage probability for different design terrestrial BS density  $\lambda_\mathrm{b}$. It is clear from the figure that when we have low BS density, the satellite network takes over and dominates the hybrid service and the coverage probability converges to that of the satellite coverage probability. In order to visualize the impact of user's density $\lambda_\mathrm{d}$ we depict in Fig.~\ref{Fig_ProbIoT} the hybrid coverage with different satellites constellation sizes. It is clear that as the number of users increases, the coverage probability deteriorates because of the increasing interference. With the aid of densifying the satellite constellation, the coverage probability can be improved again since more serving satellites are becoming closer to the zenith, i.e. smaller average contact angle leads to enhanced received powers.
\begin{figure}
	\centering
	\includegraphics[width=\linewidth]{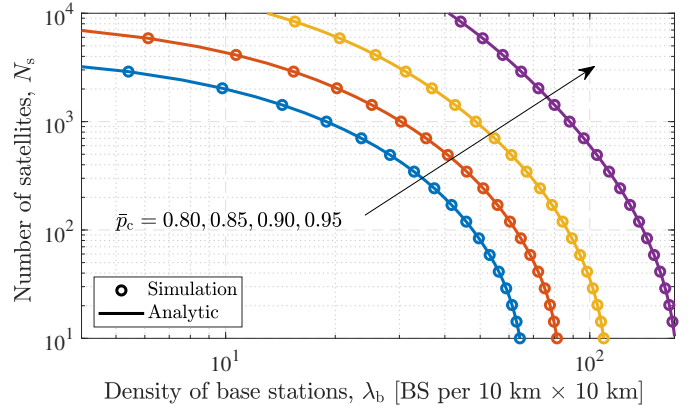}
	\caption{Hybrid operating curves for different QoS levels.}
	\label{Fig_Lambda}
	\footnotesize
\end{figure}
\section{Network Design Parameters}
\begin{figure}
	\centering
	\includegraphics[width=\linewidth]{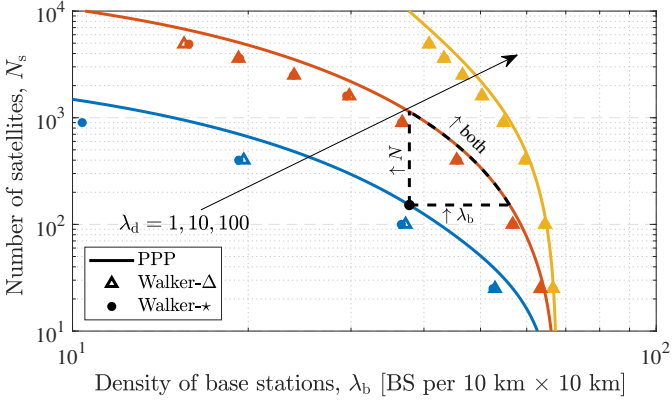}
	\caption{Hybrid operating curve at a target $\bar{p}_\mathrm{c}=0.8$ showing the impact of increasing $\lambda_{\mathrm{d}}$, comparing PPP, Walker-delta, and Walker-star constellations.}
	\label{Fig_LambdaIoT}
	\footnotesize
\end{figure}

A network design involves determining the network infrastructure parameters to maintain a target QoS, i.e. coverage probability, $\bar{p}_\mathrm{c}$, for a given user density. While a typical terrestrial network requires evaluating the BS density only, a hybrid network requires evaluating a combination of the number of satellites as well as the BS density. Thus, a designer has the choice of different combinations of $N_\mathrm{s}$ and $\lambda_\mathrm{b}$ for a given target performance. In Fig.~\ref{Fig_Lambda} we depict an example of such combinations, we call the resulting curve by \textit{hybrid network operating curve}, where each curve represents a desired QoS or $\bar{p}_\mathrm{c}$. These curves are crucial to inform strategic deployment and expansion decisions of such networks, for a given BS density, the minimum number of satellites required to maintain a certain QoS, i.e. a point on desired curve, is obtained as follows,
\begin{equation}\label{Eq_NsOptimal}
	N_{\mathrm{s}}^{\star} =\argmin_{N_\mathrm{s}}\left|\bar{p}_\mathrm{s}- \frac{\bar{p}_\mathrm{c}-\bar{p}_\mathrm{b}(\lambda_{\mathrm{b}})}{1-\bar{p}_\mathrm{b}(\lambda_{\mathrm{b}})}\right|~,
\end{equation}
where $N_{\mathrm{s}}^{\star}$ is the number of satellites required to maintain a given $\bar{p}_\mathrm{c}$ for a fixed $\lambda_{\mathrm{b}}$. On the other hand, for a fixed number of satellites, we obtain the BS density that maintains a certain $\bar{p}_\mathrm{c}$ as follows,
\begin{equation}\label{Eq_LambdabOptimal}
	\lambda_{\mathrm{b}}^{\star} =\argmin_{\lambda_{\mathrm{b}}}\left|\bar{p}_\mathrm{b}- \frac{\bar{p}_\mathrm{c}-\bar{p}_\mathrm{s}(N_{\mathrm{s}})}{1-\bar{p}_\mathrm{s}(N_{\mathrm{s}})}\right|~.
\end{equation}
The hybrid operating curve can be obtained for a QoS level by varying either $N_\mathrm{s}$ or $\lambda_\mathrm{b}$ for a fixed $\bar{p}_\mathrm{c}$. Both \eqref{Eq_NsOptimal} and \eqref{Eq_LambdabOptimal} can be evaluated by utilizing numerical optimization methods such as the Quasi-Newton algorithm~\cite{Optimization}. The coverage of PPP satellite constellations behaves as a lower bound or worst-case scenario when compared to more regular constellations such as Walker-delta (e.g. Starlink) and Walker-star (e.g. OneWeb). As such, these operating curves can provide designers with bounds for an appropriate expansion strategy for different densities of users. This can be seen in Fig.~\ref{Fig_LambdaIoT} where the operating curves are compared between the PPP, Walker-delta, and Walker-star constellations for an example height of 500~km. The number of planes for both Walker constellations is assigned to $\sqrt{N_\mathrm{s}}$ such that the number of satellites per plane is equal to the number of orbital planes. On the other hand, the inclination angles for Walker-delta and Walker-star constellations are chosen as 86.4$^\circ$ and 53$^\circ$ respectively. These operating curves provide a rapid insight into the performance of the network such that for an increase in the number of devices, one can opt to increase $N_\mathrm{s}$ only (vertical curve shift), increase $\lambda_{\mathrm{b}}$ (horizontal curve shift), or increase both (region between vertical and horizontal shifts) as illustrated by the dashed lines in Fig.~\ref{Fig_LambdaIoT}.

\section{Notes on Simulation}
A Monte Carlo simulator is developed to validate the proposed framework. It generates three independent random processes as follows; (i) a process that represents the randomly generated satellite points on a sphere with an altitude $R_{\tinyEarth} + h$, (ii) a process that represents the terrestrial BS, randomly scattered on Earth's surface, and (iii) a process that represents the users, also randomly scattered on Earth's surface. Note that the simulator assumes the Earth as an ideal sphere with an average radius of $R_{\tinyEarth}$. In order to achieve a uniform distribution of points on the spherical surface, the coordinates are generated as, $\vartheta \sim \mathcal{U}(0,2\pi)$ and $\phi \sim \asin[\mathcal{U}(-1,1)]$ where $\mathcal{U}(a,b)$ is the uniform random variable between $a$ and $b$. Table~\ref{Table_Notations} summarizes the parameters used in the simulation. The center frequency is set at 2 GHz since we follow the channel model parameters obtained based on real measurements in~\cite{9257490}. Furthermore, shared spectrum IoT is limited to low frequency bands such as the L and S bands by spectrum regulators~\cite{9205874}. At these low frequencies, we can assume that the air absorption attenuation is negligible based on the ITU model~\cite{ITU}.

\begin{table}
	\caption{Notations and Symbols}
	\centering
	\begin{tabularx}{3.49in}{l l l}
		\hline\hline \\[-1.5ex]
		Symbol & Value [Unit] &Definition \\
		\hline\\ [-1.5ex]
		$R_{\tinyEarth}$ & 6371~[km] & Earth's average radius \\
		$h$ & 500~[km] & Satellite orbit height \\
		$f$ &2~[GHz]~\cite{9257490} &Center frequency \\
		$l_{\mathrm{air}}$ &0~[dB]~\cite{ITU} &Air absorption model \\
		$\beta$ &2.3~\cite{9257490} &LoS probability parameter \\
		$\mu_{\mathrm{LoS}}$ &0~[dB]~\cite{9257490} &LoS excess path-loss mean \\
		$\sigma_{\mathrm{LoS}}$ &2.8~[dB]~\cite{9257490} &LoS excess path-loss standard deviation \\
		$\mu_{\mathrm{NLoS}}$ &12~[dB]~\cite{9257490} &NLoS excess path-loss mean \\
		$\sigma_{\mathrm{NLoS}}$ &9~[dB]~\cite{9257490} &NLoS excess path-loss standard deviation \\
		$a$ &3.68~\cite{8935360} &Path-loss exponent \\
		$P$ & 23~[dBm] & Device transmit EIRP \\
		$\psi$ & 2$\pi$~(isotropic) & Satellite beamwidth\\
		$\gamma_{\mathrm{o}}$ & -20~[dB] & Target (threshold) SINR\\
		$W_{\mathrm{b}}$ & -117~[dBm] & BS average noise power\\
		$W_{\mathrm{s}}$ & -130~[dBm] & Satellite average noise power\\
		$\kappa_{\mathrm{b}}$ & -20~[dB] & Terrestrial interference mitigation factor\\
		$\kappa_{\mathrm{s}}$ & -20~[dB] & Satellite interference mitigation factor\\
		$D$ & 1\% & Transmission duty cycle\\
		\hline 
	\end{tabularx}
	\label{Table_Notations}
\end{table}

\section{Conclusion}\label{Sec_Conc}
This paper presented an analytical approach for capturing the performance of a hybrid uplink network composing of terrestrial BS and an overlaying satellite constellation. The approach utilizes tools from stochastic geometry to capture the system as three independent random processes. The formulated framework provides designers with tools to evaluate the needed network infrastructure. The paper presents the effect of the increase of user density on the required network parameters to maintain the target quality of service allowing the projection of the future performance of the network.

\ifCLASSOPTIONcaptionsoff
\newpage
\fi
\bibliographystyle{IEEEtran}
\bibliography{HybridStochastic}

\begin{thebibliography}{10}
\providecommand{\url}[1]{#1}
\csname url@samestyle\endcsname
\providecommand{\newblock}{\relax}
\providecommand{\bibinfo}[2]{#2}
\providecommand{\BIBentrySTDinterwordspacing}{\spaceskip=0pt\relax}
\providecommand{\BIBentryALTinterwordstretchfactor}{4}
\providecommand{\BIBentryALTinterwordspacing}{\spaceskip=\fontdimen2\font plus
\BIBentryALTinterwordstretchfactor\fontdimen3\font minus
  \fontdimen4\font\relax}
\providecommand{\BIBforeignlanguage}[2]{{%
\expandafter\ifx\csname l@#1\endcsname\relax
\typeout{** WARNING: IEEEtran.bst: No hyphenation pattern has been}%
\typeout{** loaded for the language `#1'. Using the pattern for}%
\typeout{** the default language instead.}%
\else
\language=\csname l@#1\endcsname
\fi
#2}}
\providecommand{\BIBdecl}{\relax}
\BIBdecl

\bibitem{7811844}
M.~{Jia}, X.~{Gu}, Q.~{Guo}, W.~{Xiang}, and N.~{Zhang}, ``Broadband hybrid
  satellite-terrestrial communication systems based on cognitive radio toward
  {5G},'' \emph{IEEE Wireless Communications}, vol.~23, no.~6, pp. 96--106,
  2016.

\bibitem{1522108}
B.~{Evans}, M.~{Werner}, E.~{Lutz}, M.~{Bousquet}, G.~E. {Corazza}, G.~{Maral},
  and R.~{Rumeau}, ``Integration of satellite and terrestrial systems in future
  multimedia communications,'' \emph{IEEE Wireless Communications}, vol.~12,
  no.~5, pp. 72--80, 2005.

\bibitem{9313025}
A.~{Al-Hourani}, ``An analytic approach for modeling the coverage performance
  of dense satellite networks,'' \emph{IEEE Wireless Communications Letters},
  vol.~10, no.~4, pp. 897--901, 2021.

\bibitem{9079921}
N.~{Okati}, T.~{Riihonen}, D.~{Korpi}, I.~{Angervuori}, and R.~{Wichman},
  ``Downlink coverage and rate analysis of low earth orbit satellite
  constellations using stochastic geometry,'' \emph{IEEE Transactions on
  Communications}, vol.~68, no.~8, pp. 5120--5134, 2020.

\bibitem{9347980}
A.~{Yastrebova}, I.~{Angervuori}, N.~{Okati}, M.~{Vehkaperä}, M.~{Höyhtyä},
  R.~{Wichman}, and T.~{Riihonen}, ``Theoretical and simulation-based analysis
  of terrestrial interference to {LEO} satellite uplinks,'' in \emph{GLOBECOM
  2020 - 2020 IEEE Global Communications Conference}, 2020, pp. 1--6.

\bibitem{9257490}
A.~{Al-Hourani} and I.~{Guvenc}, ``On modeling satellite-to-ground path-loss in
  urban environments,'' \emph{IEEE Communications Letters}, vol.~25, no.~3, pp.
  696--700, 2021.

\bibitem{HaenggiBook}
M.~Haenggi, \emph{Stochastic Geometry for Wireless Networks}.\hskip 1em plus
  0.5em minus 0.4em\relax Cambridge University Press, 2013.

\bibitem{Optimization}
D.~F. Shanno, ``Conditioning of quasi-newton methods for function
  minimization,'' \emph{Mathematics of Computation}, vol.~24, no. 111, pp.
  647--656, 1970.

\bibitem{9205874}
B.~{Al Homssi}, A.~{Al-Hourani}, Z.~{Krusevac}, and W.~S.~T. {Rowe}, ``Machine
  learning framework for sensing and modeling interference in {IoT} frequency
  bands,'' \emph{IEEE Internet of Things Journal}, vol.~8, no.~6, pp.
  4461--4471, 2021.

\bibitem{ITU}
``{Propagation Data Required for the Design of Earth-Space Land Mobile
  Telecommunication Systems},'' International Telecommunication Union (ITU), P
  Series document ITU-R P.681-6, Tech. Rep., 2019.

\bibitem{8935360}
B.~{Al Homssi}, A.~{Al-Hourani}, S.~{Chandrasekharan}, K.~M. {Gomez}, and
  S.~{Kandeepan}, ``On the bound of energy consumption in cellular {IoT}
  networks,'' \emph{IEEE Transactions on Green Communications and Networking},
  vol.~4, no.~2, pp. 355--364, 2020.

\end{thebibliography}

\end{document}